\documentclass[aps,amsmath,amssymb,letter,scriptaddress,prl,showkeys]{revtex4}

	\usepackage{amsmath}
	\usepackage{amsthm}
	
	\usepackage{amssymb}
	\usepackage{pgfplots}
	\usepgfplotslibrary{groupplots}
	\usepackage{mathtools}
	\usepackage{makeidx}
	\usepackage{amsfonts}
	\usepackage[ansinew]{inputenc}
	\usepackage[usenames,dvipsnames]{pstricks}
	\usepackage{subfigure}
	\usepackage{epsfig}
	\usepackage{pst-grad} 
	\usepackage{pst-plot} 
	\usepackage[colorlinks,hyperindex]{hyperref}
	\makeatletter

\newcommand{\Rmnum}[1]{\expandafter\@slowromancap\romannumeral #1@}
\makeatother

\makeindex

\begin{document}

\title{Zipf's law emerges asymptotically during phase transitions in communicative systems}

\author{Bohdan B. Khomtchouk}
\email{b.khomtchouk@med.miami.edu}
\affiliation{University of Miami Miller School of Medicine \\ Center for Therapeutic Innovation and Department of Psychiatry and Behavioral Sciences \\ 1120 NW 14th Street Suite 1463, Miami, FL 33136, USA}

\author{Claes Wahlestedt}
\email{CWahlestedt@med.miami.edu}
\affiliation{University of Miami Miller School of Medicine \\ Center for Therapeutic Innovation and Department of Psychiatry and Behavioral Sciences \\ 1120 NW 14th Street Suite 1463, Miami, FL 33136, USA}


\begin{abstract}
Zipf's law predicts a power-law relationship between word rank and frequency in language communication systems, and is widely reported in texts yet remains enigmatic as to its origins. Computer simulations have shown that language communication systems emerge at an abrupt phase transition in the fidelity of mappings between symbols and objects. Since the phase transition approximates the Heaviside or step function, we show that Zipfian scaling emerges asymptotically at high rank based on the Laplace transform.  We thereby demonstrate that Zipf's law gradually emerges from the moment of phase transition in communicative systems.  We show that this power-law scaling behavior explains the emergence of natural languages at phase transitions.   We find that the emergence of Zipf's law during language communication suggests that the use of rare words in a lexicon is critical for the construction of an effective communicative system at the phase transition.
\end{abstract}


\maketitle

\section{Introduction}

The linguist George Kingsley Zipf made the observation that the frequency of a word is proportional to the inverse of the word's rank in a text. If the most common word occurs at frequency $n$, then the second most common word occurs at frequency $n/2$, the word with rank three at frequency $n/3$, etc. Generalized, Zipf's law \cite{zipf1949} states:

\begin{equation}
f \propto \frac{1}{r^\alpha}
\label{zipf}
\end{equation}
where $r$ is the word rank and $f$ the frequency in the text, and $\alpha$ is the scaling coefficient and is generally found to be near 1.0 for many of the texts examined \cite{ferrer_pnas, ferrer_2013, alday_2016, moreno_2016}.

There is, as of yet, no rigorous mathematical understanding for the origins of Zipf's law. The mathematician Benoit Mandelbrot generalized the relationship for infinite $n$ as the Zipf-Mandelbrot law \cite{mandelbrot1966, mandelbrot1983}, though this effort did not clarify the origins of the scaling and served more to increase the fit of the model to data for the lowest ranks \cite{montemurro, mit}.  More recently, simulation studies have shown that Zipf's law is found in the transition between referentially useless systems and indexical reference systems \cite{ferrer2003}.  However, a deterministic demonstration of why Zipf's law emerges spontaneously at a given threshold has not yet been given.  As such, the emergence of Zipf's law in communicative systems has lacked a rigorous explanation.

Ferrer i Cancho's research group formalized the least-effort principle as it applies to Zipf's law \cite{ferrer2003, ferrer_pnas, ferrer2007, ferrer2010} by employing a mutation-driven genetic algorithm. Here the listener and speaker have different and conflicting interests. The listener seeks to gain as much information as possible from a communicative exchange, and would benefit if there were no ambiguity between word-object mappings. This is the case in which the correlation between words and objects is highest; in information theory \cite{shannon}, this corresponds to a high mutual information, or $I(S,R)$ where $S$ represents the symbol and $R$ the referent or object. The speaker on the other hand looks to minimize his effort in communicating and would benefit from fewer words to choose from, assuming that the choice of words comes with an effort; in information theory, this is quantified using information entropy or $H(S)$.

\begin{figure}
\centerline{\includegraphics[width=\textwidth]{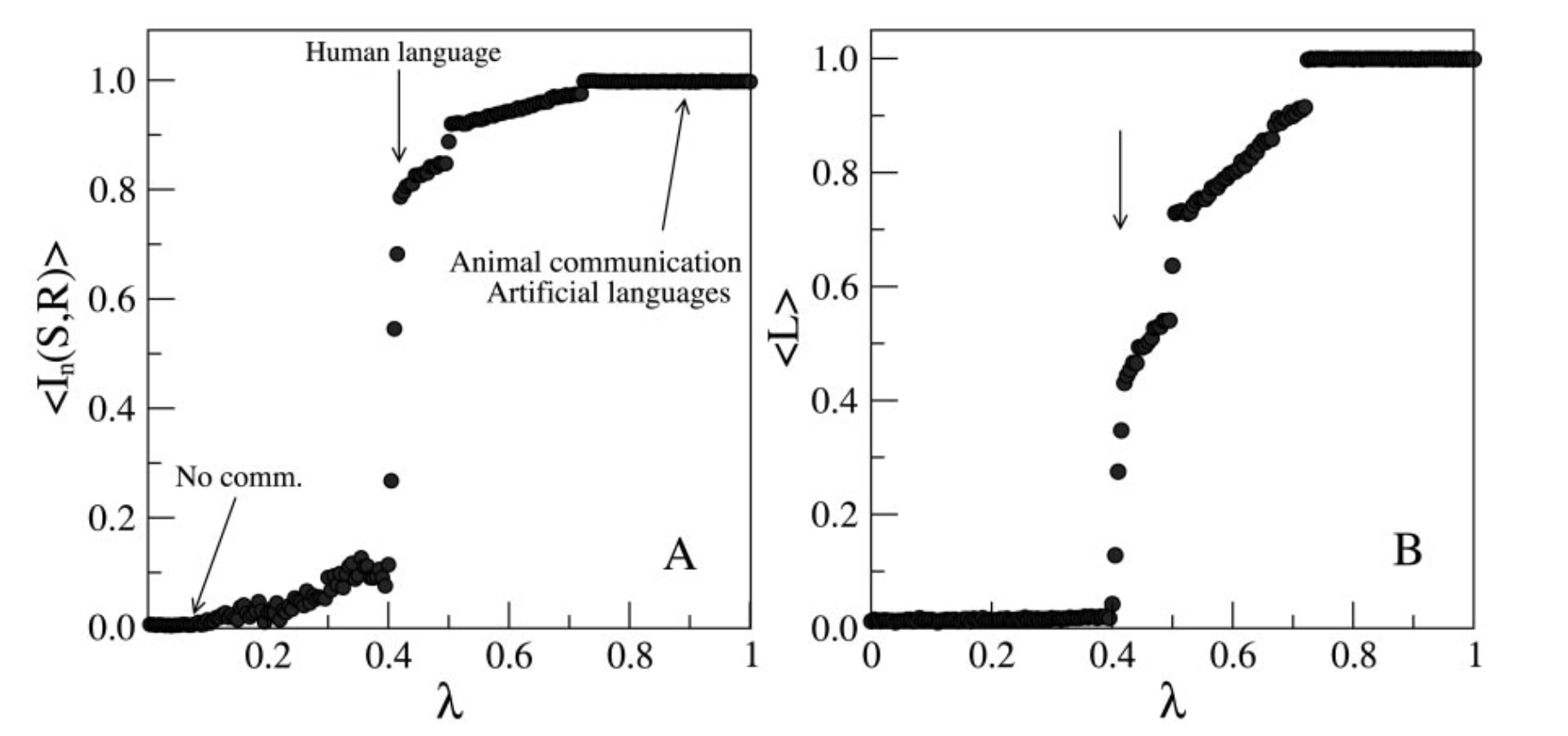}}
\caption{Phase transition in the mutual information $\langle I_n(S, R) \rangle$ and lexical size $\langle L\rangle$ of simulated languages as a function of the proportion of effort, i.e., bias ($\lambda$), devoted to listener interests as opposed to speaker interests. Reproduced with permission from Ferrer i Cancho and Sol\'e (2003).}\label{ferrer}
\end{figure}

Ferrer i Cancho \cite{ferrer_pnas} introduced an energy function based on information theory that models the speaker's and listener's interests:

\begin{equation}
\Omega(\lambda) = (1-\lambda)H(S) - \lambda I(S,R)
\label{energy}
\end{equation}
where $\lambda$ (0 $\textless$ $\lambda$ $\textless$ 1) controls the balance between the speaker interests, $H(S)$, and listener interests, $I(S,R)$.  It is found \cite{ferrer2003, ferrer_pnas, ferrer2007} that natural languages emerge at the phase transition (Fig.\,\ref{ferrer}) near $\lambda^{*} \approx 0.5$ (i.e., when listener and speaker interests are weighted equally).  For $\lambda < \lambda^{*}$, there is little or no communication because there are few words in the lexicon $\langle L\rangle$ (Fig.\,\ref{ferrer}B) while, it is assumed, the number of objects remains constant which produces tremendous ambiguity in word-meaning mappings -- one or a few words point to all the objects (i.e., low $I(S,R)$, (Fig.\,\ref{ferrer}A)).  For $\lambda > \lambda^{*}$, there is extremely efficient communication involving single word-single object mappings (i.e., high $I(S,R)$) -- though this comes at a high cost for the speaker (i.e., high $H(S)$) because the lexicon abruptly rises to the number of objects.

\begin{figure*}[ht]
\begin{center}
\centerline{\includegraphics[width=\textwidth]{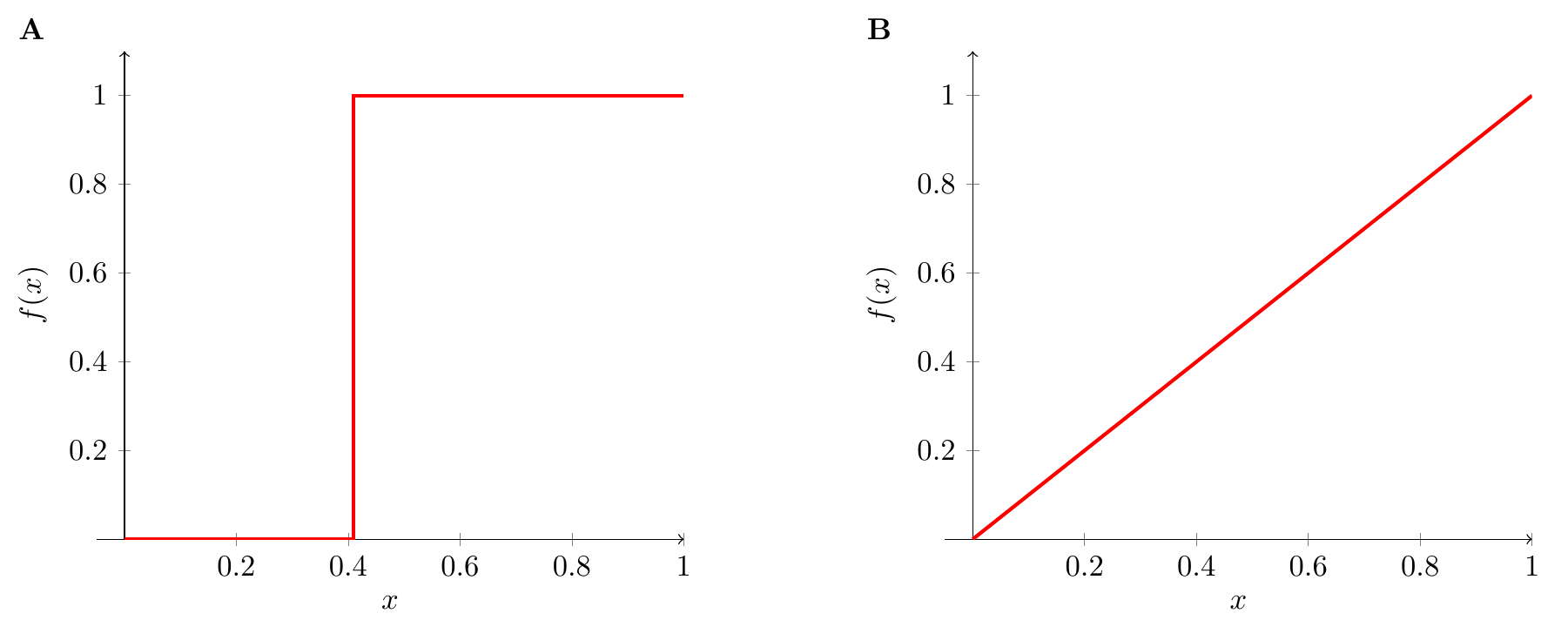}}
\caption{(A) The unit step (Heaviside function) with phase transition at $\lambda=0.41$. (B) The unit ramp function on domain $[0,1]$.}\label{afoto2}
\end{center}
\end{figure*}

The form of both of these phase transitions (Fig.\,\ref{ferrer}) lies somewhere between a step or Heaviside function and a ramp function (Fig.\,\ref{afoto2}).  The unit ramp function increases gradually, one unit per unit time.  The abrupt switching between states [$x<0$, $f(x)=0$; $x>0$, $f(x)=1$] is typical of electrical circuits \cite{spiegel} and neural systems \cite{mcculloch}.  Indeed, prior studies performed analytical derivations of global minima from equation \eqref{energy} to prove that this theoretical phase transition is well modeled by a step function \cite{ferrer2007, prokopenko}.  These studies demonstrated that the domain $\lambda < \lambda^{*}$ is characterized by single-signal systems (i.e., one signal refers to all objects), the domain $\lambda = \lambda^{*}$ is characterized by non-synonymous systems (i.e., no two signals refer to the same object, although one signal may refer to multiple objects), and the domain $\lambda > \lambda^{*}$ is characterized by one-to-one mappings between signals and objects.  

\section{Results and Discussion}
We first present some brief mathematical theory and then an exposition of the main analytical results.

\subsection{Mathematical theory}

In mathematics a transform is a method used to convert an equation in one variable to an equation in a different variable \cite{korner}. Integrals are a common type of transform and have the generalized form:

\begin{equation}
T[f(x)] = F(z) = \int_a^b \! f(x)g(x,z) \, \mathrm{d}x
\label{transform}
\end{equation}
where $f(x)$ is the function being transformed, $T$ is the generalized mathematical transform, and $g(x,z)$ is the kernel of the transform. When the definite integral is evaluated, the variable $x$ drops out of the equation and one is left with a function purely of $z$.

In a Laplace transform \cite{spiegel} ($T \equiv \mathcal{L}$), the kernel is the negative exponential $e^{-xz}$, which serves as a damping function. In the special case that $f(x)$ is the unit step function (Fig.\,\ref{afoto2}A), the Laplace transform simply yields $1/z$. For example, in electrical engineering, the Laplace transform is often used to map the behavior of functions in the time domain, $f(t)$, to the frequency domain, $F(z)$.

\subsection{Mathematical analysis}

We use the Laplace transform, $\mathcal{L}$, to map communicative functions to corresponding frequencies.  Consider the function to transform as $N(\lambda)$: the lexical size $\langle L\rangle$ of a language (i.e., the number of words in the language that are connected and have non-zero probability) as a function of the bias, $\lambda$, imparted to the listener over the speaker (Fig.\,\ref{ferrer}B). Because the lexicon size and word-meaning mappings abruptly change at the phase transition near $\lambda^{*}$ (Fig.\,\ref{ferrer}A,B), we can substitute the unit step function (Fig.\,\ref{afoto2}A) for $N(\lambda)$. The kernel $e^{-\lambda r}$ allows the mapping of word frequency from the simple transform converging to $1/r$ at increasing values of $r$:

\begin{equation}
\begin{split}
\mathcal{L}[N(\lambda)] = \int_0^1 \! N(\lambda)e^{-\lambda r} \, \mathrm{d}\lambda =\int_0^1 \! (1)e^{-\lambda r} \, \mathrm{d}\lambda = \frac{1}{r}(1-e^{-r}) = N(r)
\end{split}
\label{transform}
\end{equation}

We emphasize five key points justifying the utility of the kernel $e^{-\lambda r}$: 
\begin{itemize}
\item Lexical size of a language has been transformed from a function of bias ($\lambda$) to a function of rank ($r$).
\item $\lambda > \lambda^{*}$ (one-to-one mappings between signals and objects) corresponds to high $r$ (rare, unique signals specific to one object).  $\lambda < \lambda^{*}$ (single-signal systems where one signal refers to all objects) corresponds to low $r$ (frequent, repetitive words referring to multiple objects).
\item The y-axis is preserved under the transformation: it is still the number of words in the language (i.e., frequency).
\item Applying dimensional analysis to the Laplace transform technique validates the prerequisite of dimensionless products, as the product $-\lambda r$ is dimensionless since $\lambda$ is a constant in the range $[0,1]$ (Fig.\,\ref{ferrer}) and $r$ is a rank ($r \in \mathbb{N}$) corresponding to a specific word (i.e., signal) in the lexicon.
\item By definition, the Laplace integral transform spans 0 to $\infty$.  Since the bias, $\lambda$, imparted to the listener over the speaker has domain $[0,1]$, integration is constrained to these limits.  It can be easily verified that integrating over $[0, \infty]$, although not physically meaningful in this case, gives purely $1/r$ \cite{spiegel}.  Integration over the domain $[0,1]$ of the bias shows that $1/r$ emerges as $r \to \infty$, a hallmark of complex languages, which possess many words (where high $r$ corresponds to rare words in the lexicon).  
\end{itemize}

\begin{figure*}[ht]
\begin{center}
\centerline{\includegraphics[scale = 0.7]{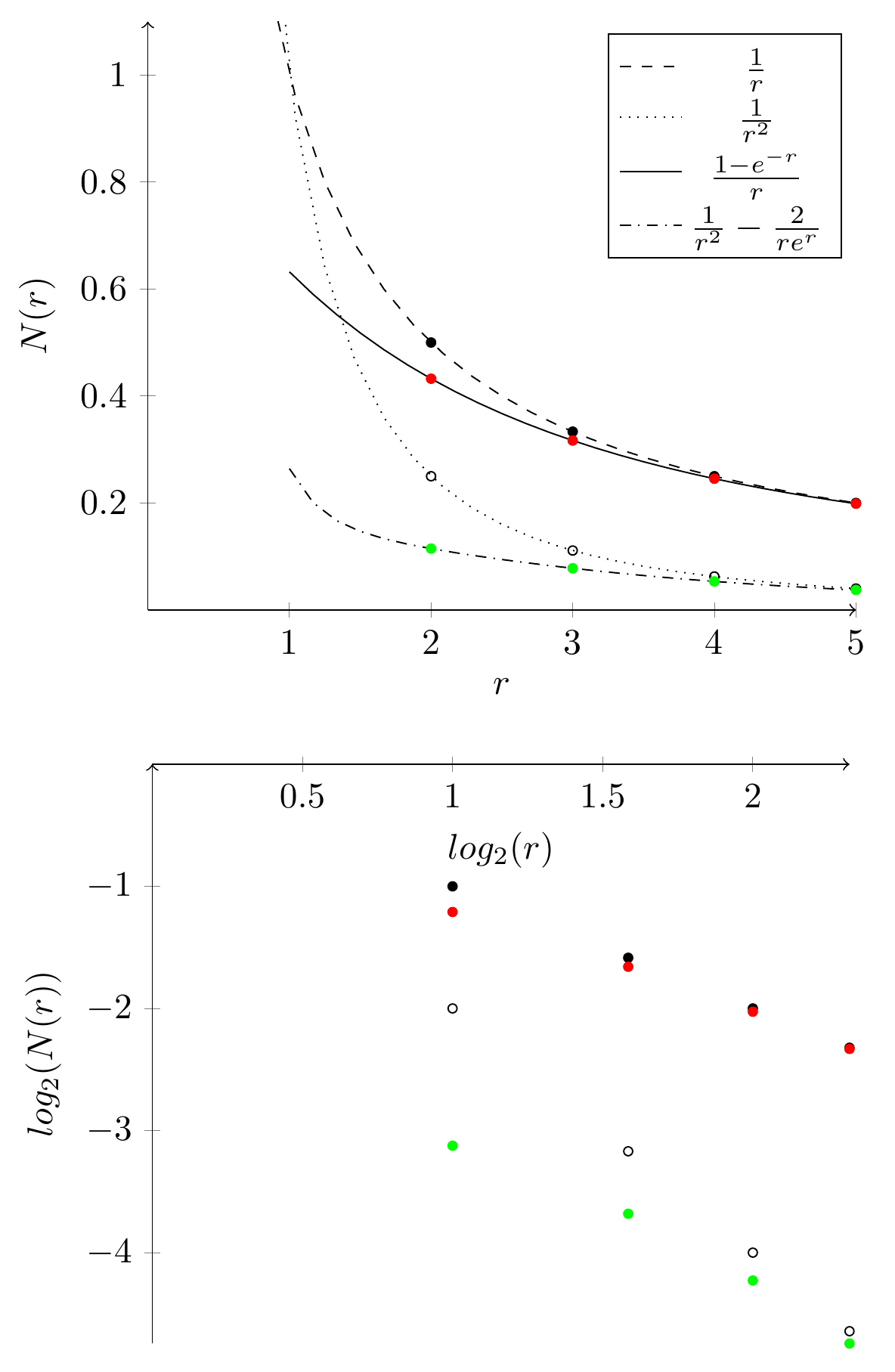}}
\caption{Zipf's law is approached asymptotically quite quickly (e.g., at 5+ words in the lexicon).  Solid dots emphasize the discrete nature of the rank variable, where it is assumed apriori that a language is defined by the existence of at least two words in the lexicon.  $N(r)$ is the lexical size of the language as a function of the rank $r$.  log-log scale is shown for plot comparison.}\label{afoto3}
\end{center}
\end{figure*}

The Laplace transform shows that Zipf's law emerges deterministically at the phase transition, for increasing values of $r$, where the $r$ term is exponentiated by 1. This mathematical operation shows that there is a connection between the rank of the $r\textsuperscript{th}$ word and its frequency in the lexicon, $N(r)$, provided the language is organizing around a phase transition in mutual information and lexicon size.  

Likewise, investigating the other boundary (Fig.\,\ref{afoto2}B), we can substitute the unit ramp function for $N(\lambda)$ and perform the Laplace transform, ultimately yielding $1/r^2$ for $r \to \infty$:

\begin{equation}
\begin{split}
\mathcal{L}[N(\lambda)] = \int_0^1 \! N(\lambda)e^{-\lambda r} \, \mathrm{d}\lambda =\int_0^1 \! (\lambda)e^{-\lambda r} \, \mathrm{d}\lambda = -\frac{2}{re^{r}} + \frac{1}{r^2} = N(r)
\end{split}
\label{transform2}
\end{equation}

Thus, depending on how abrupt the phase transition is, one should expect most words in a complex language (i.e., a language where $r \to \infty$ and high $r$ corresponds to rare words in the lexicon) to scale within the range:

\begin{equation}
\frac{1}{r^2} \leq N(r) \leq \frac{1}{r}
\label{inequality}
\end{equation}
or, in terms of the Zipfian exponent, $1 \leq \alpha \leq 2$, which is typically found to be the case \cite{ferrer_last, moreno_2016}.  Since there are infinitely many points within the boundaries of this $\alpha$ domain, this supports the notion \cite{dickman} that Zipf's law (i.e., $\alpha \approx 1$ cases) is found only in a vanishing fraction of the total minimum-cost solutions at $\lambda = 1/2$, i.e., any reasonably sized Zipf distribution has essentially zero probability of appearing in the set of minimum-cost matrices at $\lambda = 1/2$.  

Nevertheless, assuming that a language is defined by a lexicon composed of at least two unique words, Zipf's law is rapidly approached in an asymptotic fashion for lexicons composed of five or more words (Fig.\,\ref{afoto3}).  However, from a theoretical standpoint, $\alpha = 1$ can only be achieved at $r = \infty$.  Yet from a practical standpoint, it is often enough to attain a Zipfian minimum-cost state at the phase transition by utilizing a small but specific (i.e., rare and non-synonymous) lexicon that possesses one-to-one mappings between signals and objects.  We demonstrate this event by constructing a recommender system based on these principles.

\subsection{Practical demonstration}

We wish to show that at least five unique words, on average, are required for a language communication system to achieve a phase transition.  Recommender systems are the quintessential example of a language communication system: after some specific amount of training data has been inputted, recommender systems go from useless to helpful, thereby achieving a phase transition in communication, where the speaker is the human and the listener is the computer.  Current work is underway to develop and test such artificial intelligence systems based on principles in the field of natural language processing.    

\section{Conclusions}

We show that the Laplace transform maps communicative functions of speaker-listener bias directly to ranks, thereby offering a deterministic explanation for the origins of Zipf's law.  Specifically, we demonstrate that for words of high rank $r$ (i.e., rare, unique, non-synonymous words), Zipf's law is asymptotically approached in the limit as $r \to \infty$ and that this is a deterministic phenomenon.  The emergence of Zipf's law during language communication suggests that the use of rare words in a lexicon is critical for the construction of an effective communicative system. Our findings thereby support existing evidence for the emergence of natural languages at phase transition points.  We demonstrate that the complexity of a language is fundamentally determined by the number of hapax legomena, which naturally leads to the abundance of rare words.  These findings on the mathematical basis of Zipf's law highlight the importance of integral transform theory to understanding information-theoretic models of communicative systems.

\section{Acknowledgments}
\begin{acknowledgments}
BBK and CW wish to thank Ramon Ferrer i Cancho for critical review of the manuscript.  BBK wishes to acknowledge the financial support of the United States Department of Defense (DoD) through the National Defense Science and Engineering Graduate Fellowship (NDSEG) Program: this research was conducted with Government support under and awarded by DoD, Army Research Office (ARO), National Defense Science and Engineering Graduate (NDSEG) Fellowship, 32 CFR 168a.
\end{acknowledgments}

\end{document}